\documentclass[twocolumn,showpacs,preprintnumbers,amsmath,amssymb]{revtex4}
\usepackage{graphicx}% Include figure files
\usepackage{dcolumn}% Align table columns on decimal point
\usepackage{bm}% bold math

\begin{document}

\preprint{/}

\title{Experimental study of short range interactions \\ in vehicular traffic}% Force line breaks with \\

\author{C\'ecile Appert-Rolland}
\affiliation{%
Laboratoire de Physique Th\'eorique, CNRS UMR 8627, Universit\'e Paris--Sud 11, B\^at 210,
91405 ORSAY Cedex, France
}%
\email{Cecile.Appert-Rolland@th.u-psud.fr}

\date{June 13, 2009}

\begin{abstract}
Single vehicle data obtained from magnetic loops
on an expressway allowed us to measure
velocity correlations and velocity differences between non--neighbor
vehicles on the same lane, showing some strong correlations even
for platoons of 7 or 8 vehicles.
The corresponding time headway distribution for non--neighbor
vehicles is also presented.
We were also able to get some informations
about inter--lane structures, through crossed time headway distributions.
It should be noticed that these last results on crossed time headways are meaningful
only because the subset of data we used corresponds to a unique stationary traffic state
--- and still contains a large amount of data, with a large fraction
of short time headways.
The link with passing maneuvers is discussed.
\end{abstract}

%\keywords{highway traffic; time headway; platoons; velocity correlations}%Use showkeys class option if keyword
\maketitle

\section{\label{intro}Introduction}

The number of vehicles present on the road network has kept
increasing in the past decades.  Technology advances 
also modify the way of driving. More and more interest
is devoted to traffic science, and physicists are more and more
involved in the field, 
especially at the connection between experimental measurements and
modelling. This interest is in particular motivated by the
development of new or more efficient measurement devices, and
by increased data storage capacities, which allow to have
a more thorough insight into the structure of traffic.

At the microscopic level, while pair longitudinal interactions are now quite well--known
\cite{knospe02b,kesting_t08,wang08,ossen_h05,brockfeld_k_w04a,duret_b_c07,punzo_s05},
there is still much to learn
about lateral interactions \cite{lubashevsky02,hidas05,toledo_z07}, and also about collective
effects --- or correlations between vehicles,
to state it otherwise \cite{neubert99a,kuhne02,gurusinghe02,hoogendoorn_o06,treiber_k_h06b}.
This knowledge would be useful though, both in the perspective
of modelling, and of security improvement. 

This paper focuses on short range interactions between vehicles,
through the analysis of empirical single vehicle data
collected
by magnetic loops
on a two--lane expressway (N118) in the suburb of Paris (France).

In a first stage, we shall 
study intra--lane velocity correlations and relate them to
time headways, showing some strong collective effects.
Actually, velocity correlations between distant vehicles
had already been measured, as a function of the number of
vehicles placed in-between \cite{neubert99a}.
Here we obtain a much more precise measurement: for a given
number of in-between cars, we obtain the full curve giving
the velocity correlation coefficient as a function of the
time-headway.
This measurement requires a much larger data set in order
to achieve sufficient precision.
Note that in the whole paper, time headways are measured
between the rear of the leading vehicle, and the front of the other
vehicle.

While the time-headway distribution between successive vehicles
has already been measured for example in
\cite{knospe02b,wang08,krbalek_s_w01,helbing01b,treiber_h03,kerner06,krbalek07},
we shall plot it here for non--neighbor vehicles.

Then results about inter--lane structures will be presented,
with a focus on inter-lane time-headway distributions,
 and their link with passing maneuvers
will be discussed.
To our knowledge, this is the first measurement of this kind.

In order to perform these measurements, our data set had
to fulfil several constraints.

First, 
as we wanted to compare several inter-lane time-headway
distributions, a necessary condition for our conclusions
to be meaningful was that these distributions would be measured
for a unique stationnary traffic state.

Second, as our interest was in short range interactions, we wanted the
fraction of short time headways in this selected traffic state to be large.

Eventually, as the above measurements require to select
events with rather restrictive criteria, a large amount of data
had to be retained in order to have good statistics.

These goals were achieved by selecting the 11am--4pm time window on a set
of 46 week days.
Indeed, the expressway 
is mainly used for daily commuting.
On week days, while severe congestion occurs in the early morning and late afternoon,
a quite homogeneous traffic state takes
place from 11am to 4pm.
On the fundamental diagram, this homogeneous traffic state is located on the upper
part of the free flow
branch, in the region where it starts to bend, in particular
because of the interactions with slower vehicles.
Indeed, the pourcentage of cars is respectively 85\% (91\%)
on the right (left) lane. The remaining of the vehicles is mainly
composed of trucks on the right lane, and of cars towing caravans or
light trailers on the left lane.
During the period of time
from 11am to 4pm, some strong short range interactions are present;
the fractions of vehicles which have
a time headway below 0.5s (resp. 1s) are 2.4\% (resp. 15\%) on the right lane,
and 4.2\% (resp. 21.5\%) on the left lane.
The average velocity is around 90 km/h on the right lane and 110 km/h on the left lane
(the legal speed limit is 110 km/h).
Eventually we obtained
a data collection with 266 173 vehicles on the right lane and
178 546 vehicles on the left lane --- an amount of data which allows for
good statistics.
Data include, for each vehicle, its passage time, velocity, length,
and some classification into cars, trucks, etc...
Passage time refers to the passage of the front
of the vehicle on the captor.

Vehicles on one given lane
are numbered according to their passage order on the loop.
This number will be called the rank of the vehicle in the
remaining of the paper.

Throughout the paper, error bars were obtained by dividing the
data into 5 subsets, computing the quantity of interest for each
subset, and then the root mean square resulting from the
five values. Error bars correspond to 2 root mean squares.

\section{\label{intra}Intra--lane velocity correlations}

The correlation coefficient between two variables $v$
and {\bf $\tilde{v}$} is defined by
\begin{equation}
c.c. = \frac{\langle v \tilde{v} \rangle -
 \langle v \rangle \langle \tilde{v} \rangle }
{\sqrt{\left[\langle v^2\rangle - \langle v\rangle^2 \right]
\left[\langle \tilde{v}^2\rangle - \langle \tilde{v}\rangle^2 \right]}}
\end{equation}
This coefficient should be $1$
(resp. $0$) for fully correlated (resp. uncorrelated) variables.

\begin{figure}
\includegraphics[width=0.9\columnwidth]{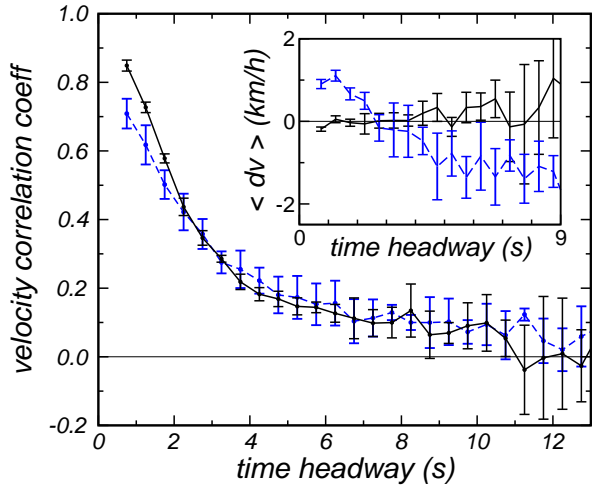}
\caption{\label{correlv1}
(Color online) For successive vehicles:
Velocity correlation coefficient and
(inset) mean velocity difference as a function of the time
headway, on the right--slow lane (black solid line) and on the left--fast
lane (grey --- blue online --- dashed line).}
\end{figure}

First we have measured 
the velocity correlation coefficient
for pairs of successive vehicles 
which have a time headway
within the interval $[k \times 0.5, (k+1)
\times 0.5[$ seconds.
The results for $1 \le k \le 25$, presented in Fig.\ \ref{correlv1},
show that short time headways induce strong
velocity correlations --- which is of course necessary in order to
avoid accidents. An exponential fit $a_0 \exp(-t_h/t_0)+a_2$ gives
a correlation time scale $t_0 = 2.3s$ on the left lane
and around 2s on the right lane.
A similar exponential relaxation is observed for
the standard deviation of the velocity difference
between successive vehicles.

As seen on Fig.\ \ref{correlv1} (inset),
the velocity difference between successive vehicles 
has its average going
to zero on the right lane, while on the left lane,
it becomes strictly positive (around 1km/h)
for short time headways, as if vehicles separated by short headways were
trying to increase their distance.
The opposite phenomenon seems to occur for larger time headways,
as if for $t_h > 4s$, vehicles were trying to catch up with
their predecessors.

\begin{figure}
\centerline{
\includegraphics[width=0.9\columnwidth]{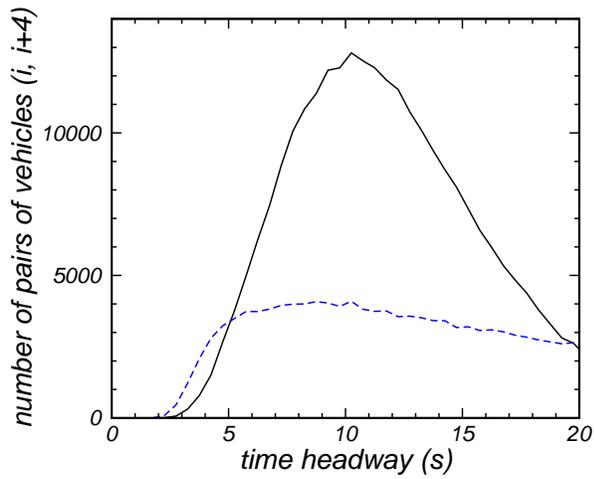}
}
\caption{\label{tiv4}
(Color online)
Distribution of time headways between vehicles of rank i and i+4,
on the right--slow lane (black solid line) and on the left--fast
lane (grey --- blue online --- dashed line).}
\end{figure}

Instead of considering successive vehicles, we shall now consider pairs
of vehicles separated by a fixed number $n$ of other vehicles and repeat
the same procedure.
For example, in the case $n=3$,
the time headway is defined as the passage time difference between
the rear of the vehicle of rank i, and the front of the vehicle of
rank i+4.
While the maximum of the time headway distribution is around 10s,
quite short time headways (below 3s) can still be observed
(figure \ref{tiv4}).
The shortest time-headways are observed on the left lane,
due to more aggressive driving.

\begin{figure}
\centerline{
\includegraphics[width=0.9\columnwidth]{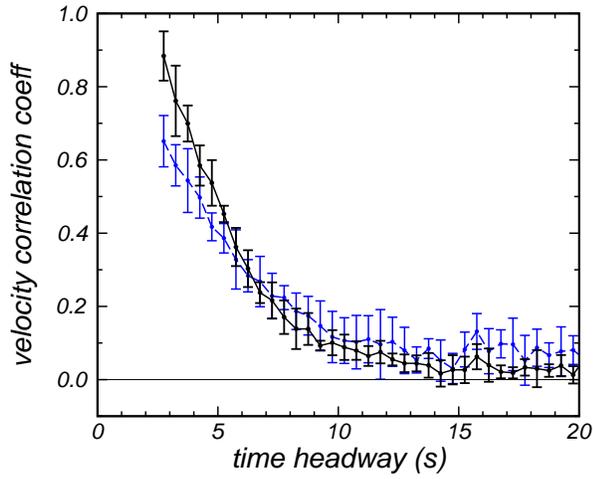}
}
\caption{\label{correlv4}
(Color online)
For pairs of vehicles of rank i and i+4:
Velocity correlation coefficient as a function of the time
headway, on the right--slow lane (black solid line) and on the left--fast
lane (grey --- blue online --- dashed line).}
\end{figure}

\begin{figure}
\centerline{
\includegraphics[width=0.9\columnwidth]{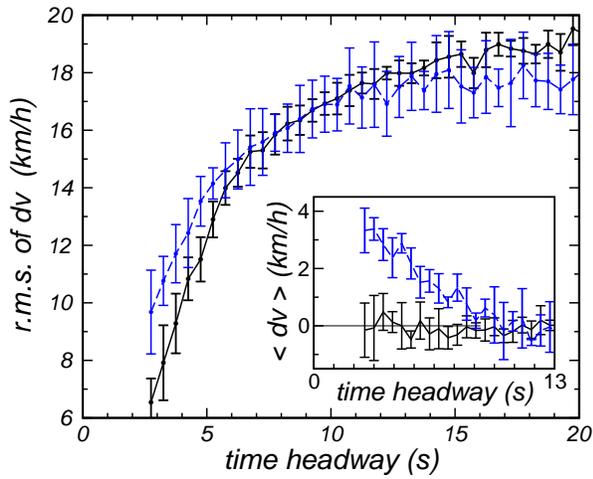}
}
\caption{\label{correldv4}
(Color online)
For pairs of vehicles of rank i and i+4:
Standard deviation and (inset) mean of the velocity difference, as a function of the time
headway, on the right--slow lane (black solid line) and on the left--fast
lane (grey --- blue online --- dashed line).}
\end{figure}

In spite of the distance between the vehicles, velocity correlations
are still very strong at short time headways,
especially on the right lane (see Fig.\ \ref{correlv4}).
The characteristic time scale obtained from an exponential fit
is now 3.6s on the left lane and 3.1s on the right lane,
to be compared with the 3.5s and 3.3s obtained from the fit
of the velocity difference standard deviation (Fig.\ \ref{correldv4}).
These results can be compared with those obtained for successive
vehicles. We notice that, if we consider the velocity correlation of
pairs of vehicles separated by a given time headway, 
it is higher when there are more cars between the two vehicles of the pair.
Or, to say it otherwise, the data support an increasing velocity
correlation with the increase of n (for a fixed time headway).
In a certain way, the in--between vehicles rigidify the binding between end
vehicles. We could also say that they transmit information
from one end car to the other.

The mean velocity difference at short time headways on the
left lane (fig.\ \ref{correldv4}, inset) is now equal to about 3 or 4 times its value for successive
cars. It is an additive phenomenon, each car trying to drive
about 1 km/h more slowly than its predecessor.

We observe that platoons of vehicles on the right lane are formed
behind a truck in about one third of the cases, while situations
are much more varied on the left lane.

Velocity correlation coefficients as large as 0.6 can still be
observed between vehicles of ranks $i$ and $i+7$ separated by a time headway
of about 6 or 7s (when the rank difference is increased further, results
are too noisy to conclude).
Thus quite long strongly correlated platoons can be observed 
in the flow.

\section{\label{inter}Across lane time headway distributions}

One important question when short time headways are observed
is whether those are stable or only transients. A situation
where one expects transient short time headways is when a
vehicle prepares itself to pass another vehicle.
We thus wondered whether vehicles that have a short time headway
on the right lane indeed have an opportunity to change lane,
i.e. whether there is a ``hole'' on their left.

\begin{figure}
\centerline{(a) 
\hskip 0.5cm
\includegraphics[width=0.7\columnwidth]{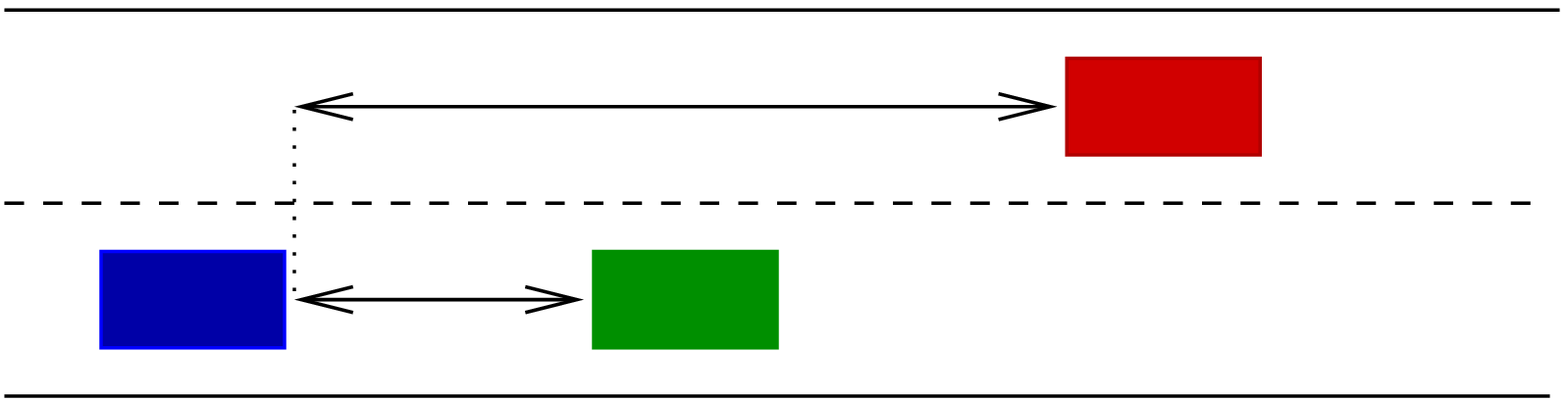}
}
\centerline{(b) 
\hskip 0.5cm
\includegraphics[width=0.7\columnwidth]{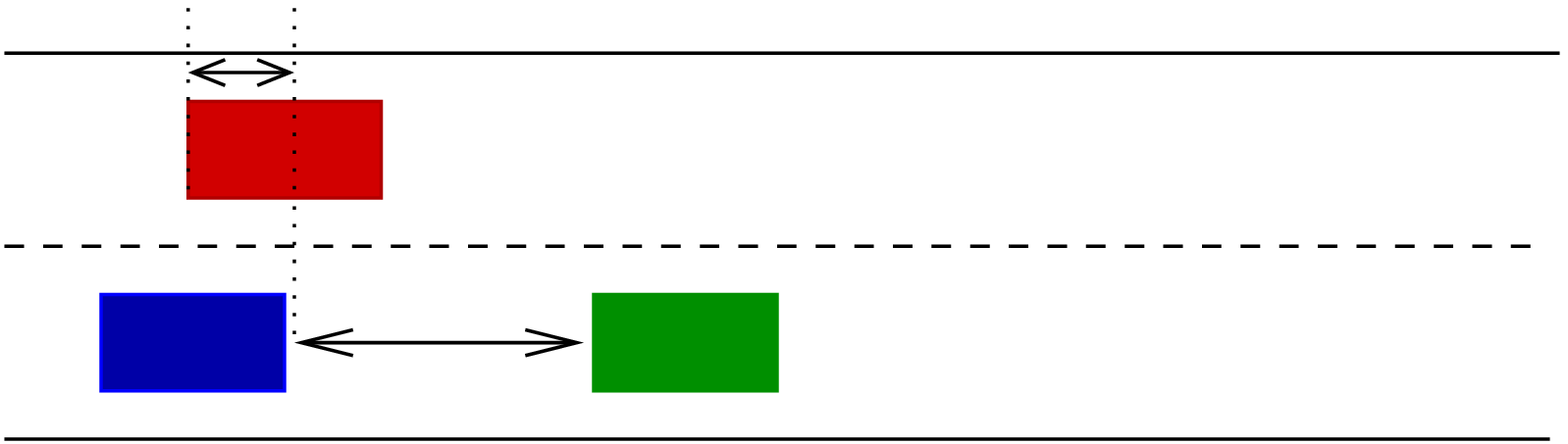}
}
\centerline{(c) 
\hskip 0.5cm
\includegraphics[width=0.7\columnwidth]{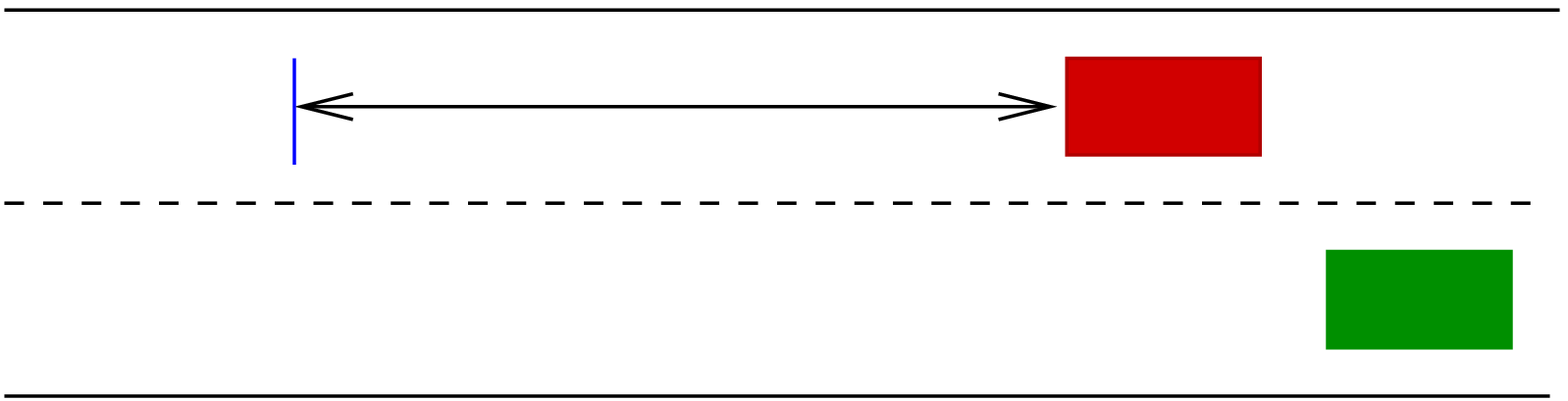}
}
\centerline{(d) 
\hskip 0.5cm
\includegraphics[width=0.7\columnwidth]{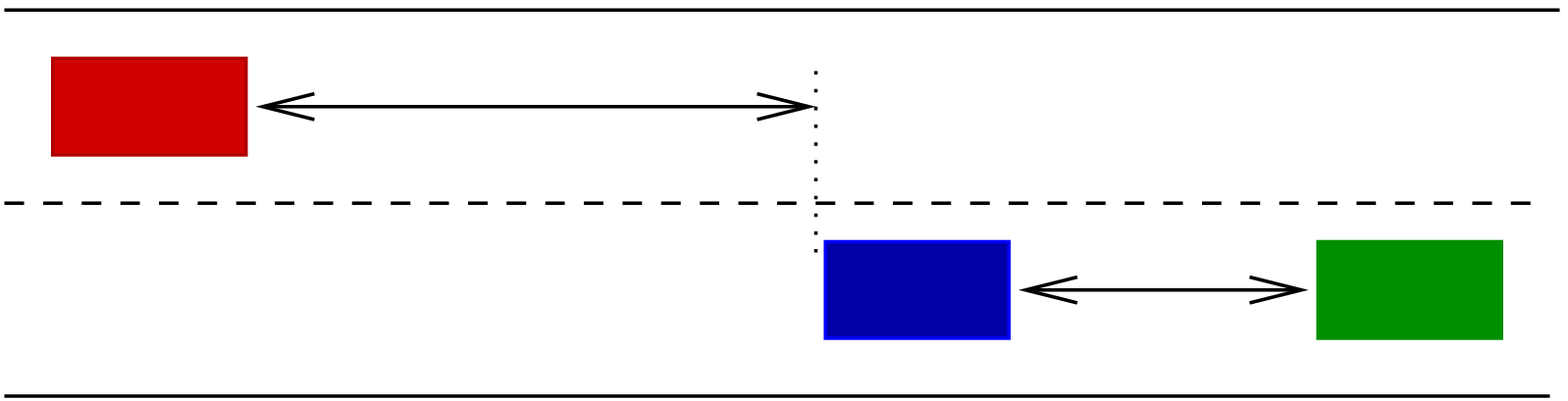}
}
\centerline{(e) 
\hskip 0.5cm
\includegraphics[width=0.7\columnwidth]{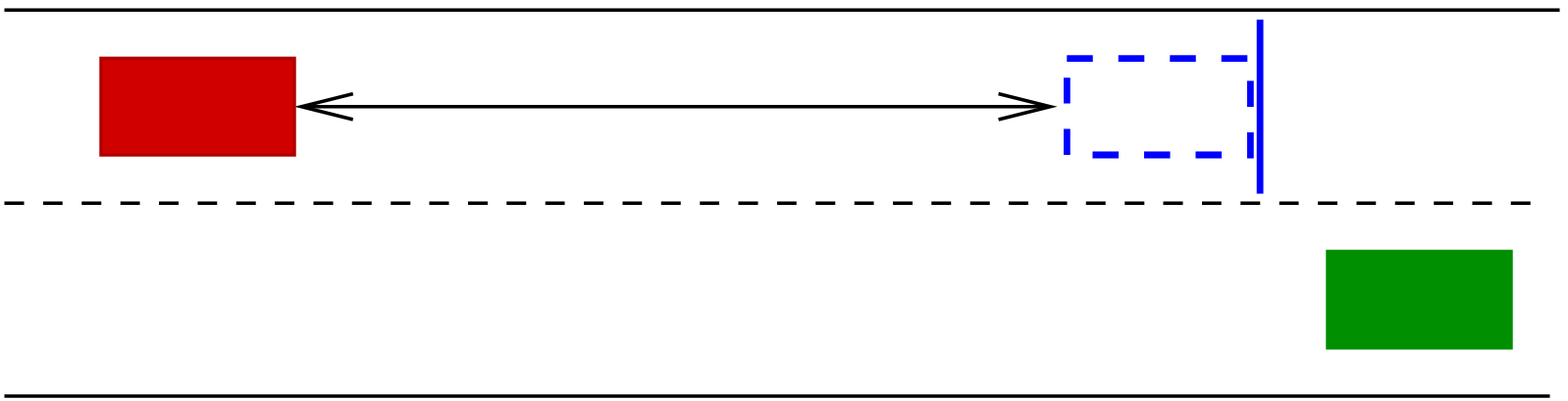}
}
\caption{
\label{schemas}
(Color online)
Schematic representation of the inter--lane time headway measurements,
for a two--lane road.
Vehicles are driving to the right.
In Fig. (a), a time headway shorter than 0.5s separates the two vehicles
on the right lane. Then, the time headway between the right lane
following car and its predecessor on the other lane is computed as
indicated by the arrow (in all the schemes, a spatial representation has been substituted
to the temporal one, though in practice, one computes the passage time difference
between the rear of the left lane vehicle
and the front of the right lane vehicle).
Fig. (b) illustrates how a negative time headway can be obtained : the front of
the left lane vehicle went first on the magnetic loop (thus it is the leader of
the right lane vehicle), but its rear arrived on the loop after the front of the right lane
vehicle.
Fig. (c), the time headway is computed between a virtual vehicle whose location
is randomly chosen, and its predecessor on the same (left) lane.
The procedure is independent of the right lane state.
Fig. (d): same as (a), 
but it is the {\em successor}
that is considered on the left--lane.
Fig. (e): same as (c), but the time headway is computed with
the successor of the virtual vehicle.
The ratio length/velocity of the virtual vehicle is chosen randomly
according to the distribution of length/velocity ratios of the right-lane
vehicles.
}
\end{figure}

We measured their time headway with the preceding vehicle
located, not on the same lane, but on the left lane, as this
is illustrated schematically on figure \ref{schemas}a.
As the preceding vehicle on the left lane may partially be
side by side with the right lane vehicle (figure \ref{schemas}b),
negative time headways can be measured.
The resulting distribution must be compared with
the distribution obtained by
the random procedure described in figure \ref{schemas}c.
Note that, as the random passage time is chosen without any exclusion rule,
the virtual vehicle and its (real) leader may partially overlap,
thus negative time headways may again be obtained.

\begin{figure}
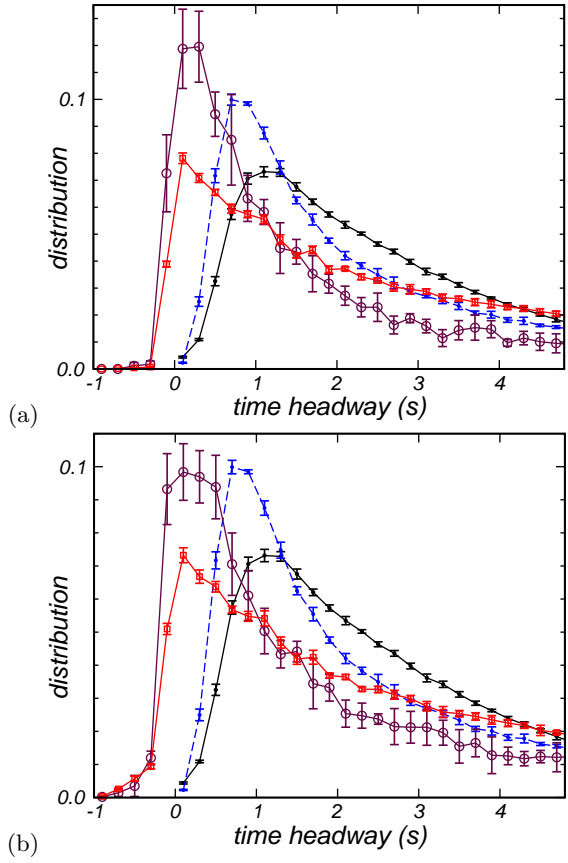

\centerline{(a) 
\includegraphics[width=0.8\columnwidth]{fig6a.eps}
}
\centerline{(b) 
\includegraphics[width=0.8\columnwidth]{fig6b.eps}
}
\caption{
\label{distf}
(Color online)
On both figures:
intra right and left lane time headway distributions (resp. solid
black and dashed grey --- blue online --- lines).
Fig.(a): distributions of the time headway with the left--lane predecessor,
 for (maroon online --- circles) the inter--lane procedure 
of fig.\ \protect{\ref{schemas}}a, or 
for (red online --- squares) the random procedure
of fig.\ \protect{\ref{schemas}}c.
Fig.(b): same as fig. (a), but it is now the {\em successor}
that is considered on the left--lane.
The random procedure (red online --- squares) corresponds
now to the procedure of fig.\ \protect{\ref{schemas}}e.
}
\end{figure}

If one tries to guess the shape of the distribution resulting from this random
procedure, two opposite effects can be expected:
on the one hand, it is more likely that the random passage time will
be chosen inside a {\em large} time interval separating the passage
of two successive vehicles. Thus more importance should be given to
large time headways.
On the other hand, the random procedure ``cuts'' into two parts the time headways,
thus the weight of small time headways should also be enhanced.
Actually, both effects are observed. As seen on figure \ref{distf}a,
if one compares the distribution obtained by the random procedure
with the intra left lane time headway distribution, 
the maximum of the distribution is shifted towards small (even negative)
time headways, while the tail at large time headways is less damped
than for the intra lane distribution.

Now, when a short time headway (less than 0.5s)
is observed on the right lane, the time headway with the vehicle on the other
(left) lane is likely to be much shorter than for a random procedure.
Thus right lane vehicles with a short headway are more likely to have
a very near neighbor ahead (or aside) in the other lane.

At this stage, this does not necessarily mean that the right lane vehicle does not intend
to overtake, it could overtake just after its left leader. In order to explore
further the possibilities that the right vehicle has of overtaking,
we repeated the same procedure for the distribution of the time headways
between a right lane vehicle with a short
time headway, and its {\em follower} on the left lane
(see a schematic representation in figure \ref{schemas}d).
The corresponding random procedure described in fig. \ref{schemas}e
is also performed.

Note that the random procedures in figures \ref{schemas}c and \ref{schemas}e
are not completely equivalent.
In order to compute the time headway as we define it, the ratio length/velocity
of the leading vehicle has to be removed from the passage time difference
in order to account for the size of the leading vehicle.
In the random procedure of fig. \ref{schemas}c, this ratio
refers to the predecessor on the left lane, while in the second case
the ratio refers to the virtual vehicle.
As a consequence, the distributions obtained from the random procedures
in figures \ref{distf}a and \ref{distf}b differ slightly.

The conclusions, drawn from figure \ref{distf}b, are very similar
to those for the forward case.
Even more negative time headways are observed.
I.e. vehicles on the right lane with a short time headway are more likely
than others to have a very close follower on the other lane.
When this occurs, they cannot overtake immediately, and their short headways
are more likely related to the impossibility to overtake immediately
rather than due to the preparation of an overtaking maneuver.

We checked also that, when a left lane vehicle is partially side by side with a
short headway right lane vehicle, the time headway with its follower on the
same left lane follows a distribution which cannot be distinguished from
the non biased intra lane time headway distribution (at least at the precision that
can be reached from our data --- statistics are less precise as this event
does not occur so often).  
Anyhow this reinforces the observation that in general, the probability that a ``hole''
can be found to the left
of right lane vehicles with short headways is smaller than for 
``standard'' vehicles.

Right lane short headways thus seem to be related in many cases
to steric constraints
in the near neighborhood, and are thus likely to last at least until the
local excess of density on the other lane is resorbed.
The frustration yielding these short time headways could also favor
potentially dangerous maneuvers, such as overtaking maneuvers with
very short inter--lane time headways.

\section{\label{conclusion}Conclusion}

The distributions of time headways between
vehicles in different lanes have been measured,
in order to characterize the environment of
the right--lane vehicles which have a short
time headway.
We stress that the comparison of distributions that we did in
section \ref{inter} is valid only because we have selected a homogeneous
traffic state.
To our knowledge it is the first time that these inter--lane time headway measurements were performed.
They rule out the idea that short time headways on the right lane
could be only due to the preparation of passing maneuvers.
Steric constraints can also induce short time headways
--- and these are expected to last for a longer time,
implying some risk factor.
In further investigations, we are trying to relate inter--lane velocity
differences with the possibility or impossibility to pass due to
steric constraints.

About intra--lane measurements, we show that
pairs of vehicles separated by more than 6 other vehicles can exhibit
strong velocity correlations --- and short time headways.
This is certainly a security issue, as it is well known
that the risk increases with the number of successive short
time headways.

From the modelling point of view, note
that strong non local correlations can be obtained
with only nearest neighbor interactions.
This is indeed the case with cellular automata, which exhibit
strong --- and most of the time pathological --- non local velocity correlations,
in spite of purely nearest neighbor interactions.
On the other hand, comparisons between car following experiments
and models
seem to show that, at least in some cases, vehicles
do interact with multiple predecessors
\cite{hoogendoorn_o06}. The conditions for
multiple interactions --- and the relevance of multiple interactions
for modelling issues --- still require investigations.

More generally, closely packed multilane microscopic structures still
deserve observations and analysis in order to be better characterized and
understood, in particular because of the crucial role that they
play in security issues.

Acknowledgments:
Work realized in the frame of the collaboration contract
2006CT025
with Laboratoire Central des Ponts et Chauss\'ees.
We thank in particular B. Jacob, M. Bry, V. Dolcemascolo,
D. Daucher, A. Koita from LCPC--Paris,
and E. Violette from CETE--Normandie,
 for fruitful discussions.
Data were collected by the LROP, and we thank V. Leray
and I. Monmousseau for their help.
We thank L. Santen for his interesting comments.

\bibliographystyle{unsrt}

\end{document}